\documentclass[preprint2]{emulateapj}

\usepackage{graphics,graphicx}
\usepackage{fancyheadings}
\usepackage{color}
\usepackage{natbib}
\usepackage{adjustbox}

\newcommand{\chandra}{\textit{Chandra}}
\newcommand{\xmm}{XMM-\textit{Newton}}

\newcommand{\nustar}{\textit{NuSTAR}}

\newcommand{\oq}{OQ\,$+208$}
\def\arcsec{\hbox{$^{\prime\prime}$}}

\shorttitle{Broadband X-ray Study of \oq}
\shortauthors{Sobolewska et al.}

\begin{document}

\title{First Hard X-ray Observation of a Compact Symmetric Object: A Broadband X-ray Study of a radio galaxy \oq\ with \nustar\ and \chandra}

\author{Ma{\l}gosia Sobolewska$^1$, Aneta Siemiginowska$^1$, Matteo Guainazzi$^2$, Martin Hardcastle$^3$,
Giulia Migliori$^4$, Luisa Ostorero$^5$, \L ukasz Stawarz$^6$}

\affil{$^1$ Harvard-Smithsonian Center for Astrophysics, 60 Garden Street, Cambridge, MA 02138, USA}
\affil{$^2$ European Space Research and Technology Centre (ESA/ESTEC),
Kepleriaan 1, 2201 AZ, Noordwijk, The Netherlands}
\affil{$^3$ School of Physics, Astronomy and Mathematics, University of Hertfordshire,
College Lane, Hatfield AL10 9AB, UK}
\affil{$^4$ INAF, Istituto di Radio Astronomia di Bologna, Via P. Gobetti 101, I-40129 Bologna, Italy}
\affil{$^5$ Dipartimento di Fisica - Universit\`a degli Studi di Torino
and Istituto Nazionale di Fisica Nucleare (INFN), Via P. Giuria 1, 10125 Torino, Italy}
\affil{$^6$ Astronomical Observatory, Jagiellonian University, ul. Orla 171, 30-244 Krak\'ow, Poland}
\smallskip
\email{msobolewska@cfa.harvard.edu}

%\maketitle

\label{firstpage}

\begin{abstract}

Compact Symmetric Objects (CSOs) have been observed with \chandra\ and \xmm\
to gain insights into the initial stages of a radio source evolution and probe the
black hole activity at the time of relativistic outflow formation. However,
there have been no CSO observations to date at the hard X-ray energies ($> 10$\,keV),
impeding our ability to robustly constrain the properties of the intrinsic X-ray
emission and of the medium surrounding the young expanding jets. We present
the first hard X-ray  observation of a CSO performed with \nustar. Our target,
\oq, is detected up to 30\,keV, and thus we establish CSOs as a new class of
\nustar\ sources. We analyze the \nustar\ data jointly with our new \chandra\
and archival \xmm\ data and find that a young, $\sim 250$ years old, radio jet
spanning the length of 10\,pc coexists with cold obscuring matter, consistent
with a dusty torus, with an equivalent hydrogen column density
$N_H = 10^{23}$--10$^{24}$\,cm$^{-2}$. The primary X-ray emission is characterized
by a photon index $\Gamma \sim 1.45 $ and intrinsic 0.5--30\,keV luminosity
$L \simeq 10^{43}$\,erg\,s$^{-1}$. The results of our spectral modeling and
broad-line optical classification of the source suggest a porous structure
of the obscuring torus. Alternatively, the source may belong to the class
of optically un-obscured/X-ray obscured AGN. The observed X-ray emission is
too weak compared to that predicted by the expanding radio lobes model,
leaving an accretion disk corona or jets as the possible origins of the X-ray
emission from this young radio galaxy.

\end{abstract}

\keywords{galaxies: active --- galaxies: jets --- galaxies: evolution --- X-rays: galaxies}

\section{Introduction}
\label{sec:intro}

Theories of radio source evolution indicate that the initial expansion
results in strong interactions between the radio outflow and the interstellar
medium (ISM) within the central regions of the host galaxy. If the central
regions are dense, then such interactions will generate strong shocks impacting
the ISM, and clearing out the path for a jet (Begelman \& Cioffi 1989;
Heinz et al. 1998; Wagner \& Bicknell 2012). Additionally, models predict
that this initial evolutionary phase 
is characterized by high radio luminosity (Begelman \& Cioffi 1989)
and that the radio plasma contained in the compact lobes produces high-energy
emission via the inverse Compton process (Stawarz et al. 2008; Migliori et al. 2014),
or through strong shocks driven into the ISM by the expanding cocoon
(Reynolds et al. 2001). However, there is little observational evidence to support 
these predictions as to date only one young radio source has been detected
in the $\gamma$-ray band (Migliori et al. 2016). Furthermore, 
the quality of the X-ray data currently available
for the young radio sources is still rather poor,
as these sources are often faint and require long \chandra\ and \xmm\
integration times to collect enough photons for a meaningful spectral analysis.
Only a small number of CSOs have been studied in X-rays, yielding observed
2--10\,keV fluxes of the order of 10$^{-14}$--10$^{-13}$\,erg\,cm$^{-2}$\,s$^{-1}$,
and none at energies above 10\,keV \citep[S16 hereafter]
{siemiginowska2009,siemiginowska+2016}.

%%%%%%%%%%%% Figure 1

\begin{figure*}
\includegraphics[width=6cm]{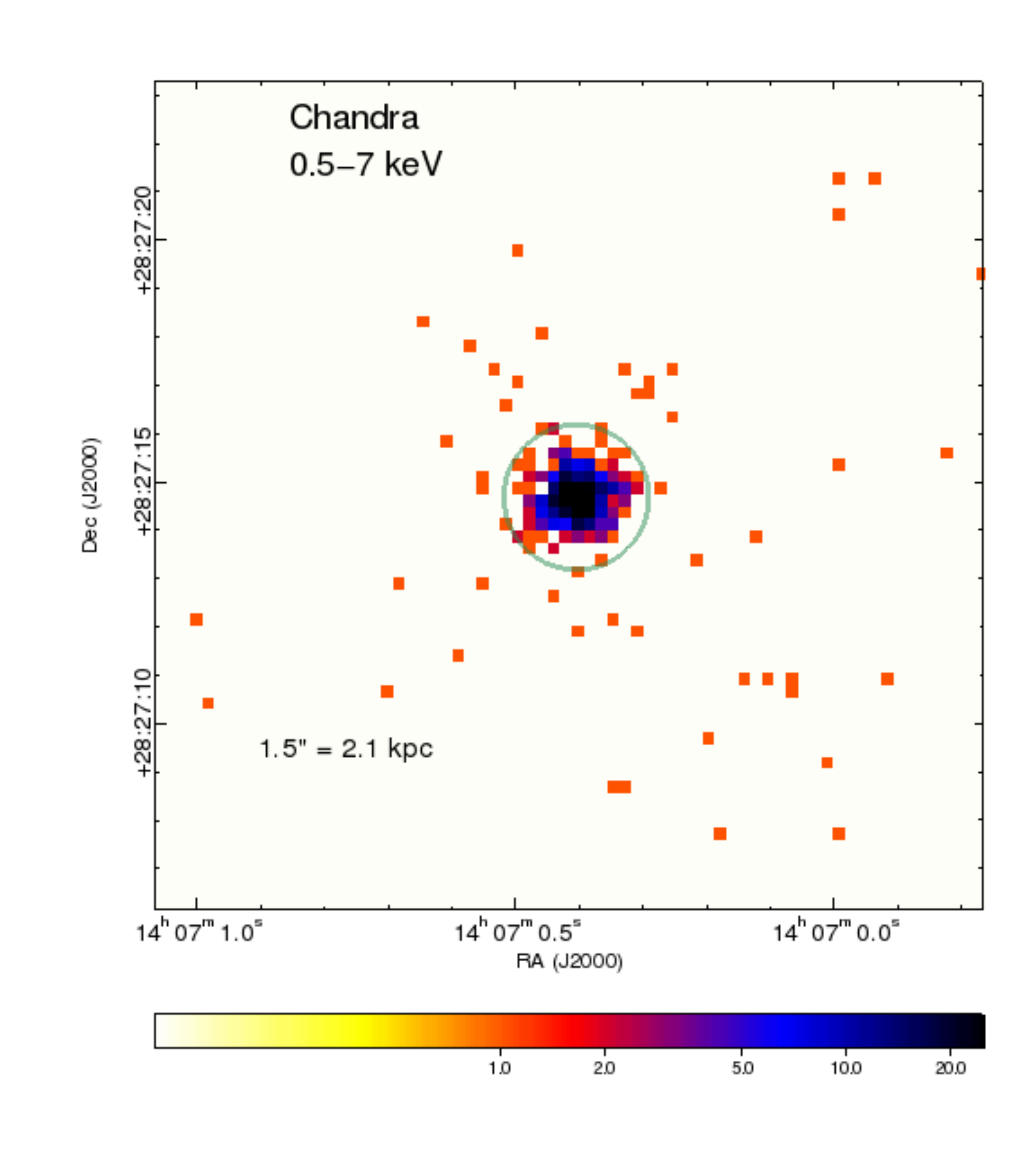}
\includegraphics[width=6cm]{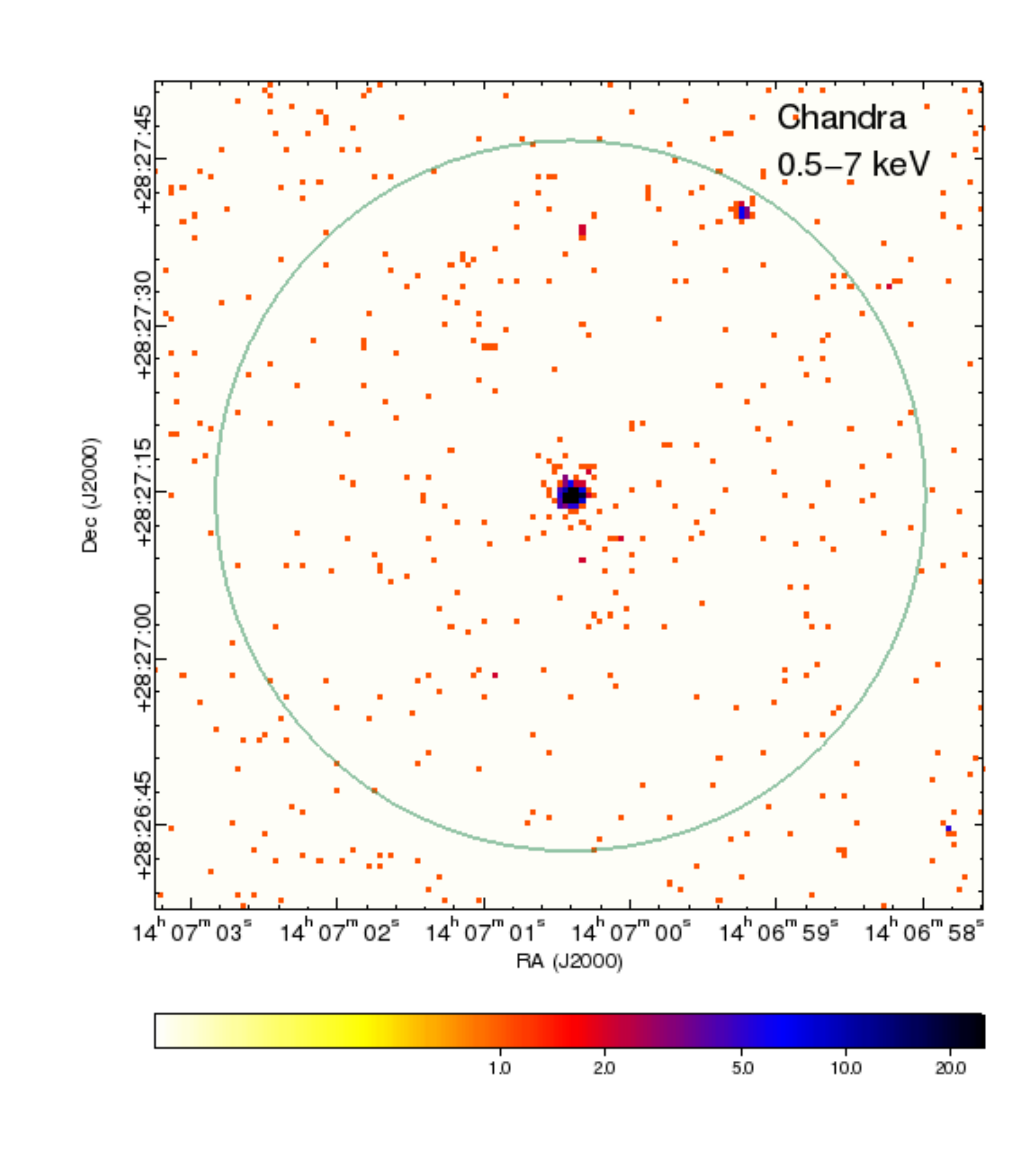}
\includegraphics[width=6cm]{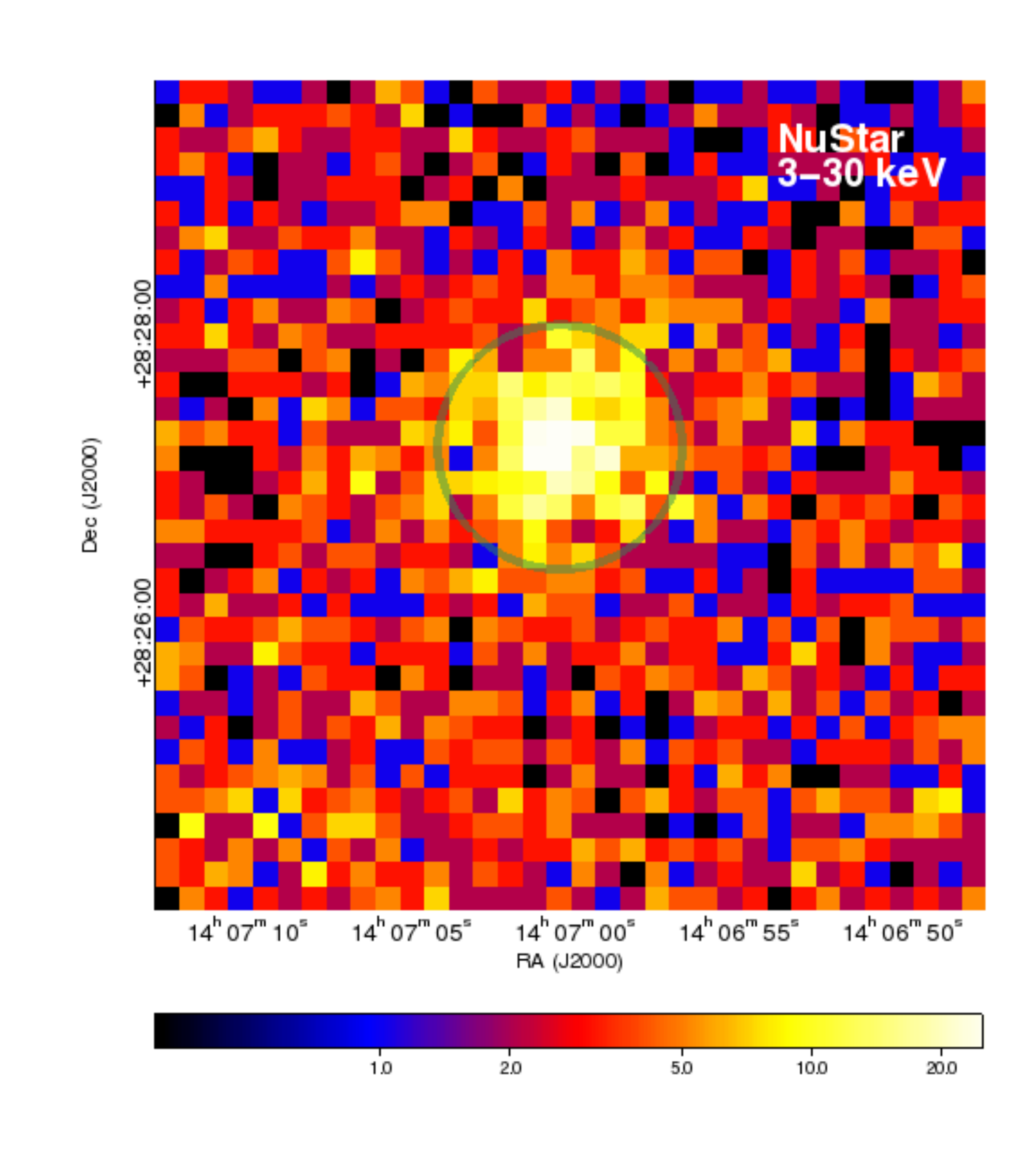}
\caption{{\bf Left:} \chandra\ ACIS-S image of \oq\ in the 0.5-7\,keV energy range, with the 
source region marked with the green circle with the radius of 1.5\arcsec\ (corresponding to 
2.1\,kpc at the redshift of the source). The pixel size is 0.249\arcsec.
{\bf Center:} The \chandra\ ACIS-S image binned to pixel size of 0.492\arcsec\ showing
a larger view of the area with the \xmm\ extraction region marked with the green circle with 
the radius of 32\arcsec. The secondary source at the $\sim$30\arcsec\ distance is visible in the
upper right section of the image.
{\bf Right:} \nustar\ FPMA 3--30\,keV image  of \oq\ with the 50\arcsec\ radii circular region 
centered on the position of \oq\ marked by the green circle. \nustar\ image is binned by a factor
of 4 with 1 pixel corresponding to $\sim$10\arcsec. The color log scale is marked on the bottom
of each image indicating the number of counts in a pixel.} 
\label{fig:fig1}
\end{figure*}

%%%%%%%%%%%%%%% Table 1

\begin{table*}
{\scriptsize
\noindent
\caption[]{\label{tab:obs} Log of X-ray observations of \oq.}
\begin{center}
\begin{tabular}{lllllcccc}
\hline\hline
& Date & Mission & Instrument & Obsid & Exposure & Total$^a$ & Net$^a$ & \\
    &      &         &			  & 	  &  [s]     & counts    & counts  &     \\
\hline
1 & 2014-09-04 & Chandra & ACIS & 16071   &   34604  & 678 & $650.0\pm26.5$ & \\
2 & 2016-06-18 & NuSTAR  & FPMA &  60201043002 & 50354 & 756 & $553.6\pm51.8$ & \\
3 & 2016-06-18 & NuSTAR  & FPMB &  60201043002 & 50150 & 704 & $484.4\pm52.8$ & \\
\hline\hline
\end{tabular}
\end{center}
\smallskip Notes:\\
$^a$ Total and background subtracted counts in a circle centered on the source position:
(1) \chandra\ with 1.5\arcsec\ radius and the energies between 0.5\,keV and 7\,keV, and
(2) \nustar\ with 49.2\arcsec radius and the energies between 3\,keV and 30\,keV.
}
\end{table*}

Nevertheless, several CSOs have recent, good signal-to-noise X-ray spectra, allowing
for detailed studies of their high-energy properties. Analysis revealed
that these sources produce complex X-ray emission in which the primary power-law continuum,
from either an accretion disk corona, jets or expanding radio lobes, is modified by
absorption and reflection processes (e.g. Guainazzi et al. 2004; Siemiginowska et al. 2008; Tengstrand et al. 2009; S16; Sobolewska et al. 2019). 

In this paper, we present an X-ray study of \oq\ (1404+286), a nearby
($z$=0.0766, \citealt{stanghellini1993}) radio source classified as a Compact Symmetric Object
(CSO) based on the properties of its radio morphology. 
The double radio structure, with two components separated by 10\,pc, the 
symmetry of the radio source, and multi-epoch radio monitoring allowed
kinematic age measurements that showed that the radio source is
255$\pm17$\,yrs old (Wu et al. 2013). The source is located in a nearby
disturbed galaxy, Mkn\,668, that shows signs of a past merger
(Stanghellini et al. 1993). Based on its optical spectrum, it is classified
as a broad-line radio galaxy (Marziani et al. 1993). The archival
\xmm\ spectrum showed a steep soft X-ray component of an unknown origin and
a hard highly absorbed X-ray continuum with a strong Fe-K$\alpha$ emission
line (Guainazzi et al. 2004). Thus, we observed the source with \chandra\
and with \nustar, in order to construct and model its broad-band X-ray
spectrum, search for any extended X-ray emission, and detect for the first
time a young radio source above 10\,keV.

Our data allow us to disentangle the individual X-ray
emission components and obtain the energetics of the primary emission associated 
with this young radio source, as well as study the properties of the environment
in which the source is expanding. Our relatively deep \chandra\ observation shows
no significant X-ray diffuse emission on the scales exceeding 3\,arcsec, indicating that
the complex emission arises from the central regions. The \nustar\ observation provides
crucial constraints on the high-energy spectral properties by extending the available 
energy coverage over which the source is detected to $\sim 30$\,keV. 
In Section 2 we present our new \chandra\ and \nustar\ data and provide the
details of the spectral model adopted in our analysis. In Section 3 we present our 
results, and we discuss them in Section 4. We summarize our findings in Section 5.

We use the most recent cosmological  parameters
(Hinshaw et al. 2013; $\rm H_0 = 69.3$\,km\,s$^{-1}$\,Mpc$^{-1}$, $\rm \Omega_m = 0.287$)
implemented as WMAP9 in the {\tt astropy.cosmology} package
(Astropy Collaboration et al. 2013).

%%%%%%%%%%%%%%% Table 2

\begin{table*}[t]
{\scriptsize
\noindent
\caption{\label{tab:models} Spectral model parameters.}
\begin{center}
\begin{tabular}{rllccc}
\hline\hline
 \# & Description & Symbol & \multicolumn{2}{c}{Value$^a$}  & Unit \\
    &     &     & $N_H$ linked & $N_H$ unlinked & \\
    & (1) & (2) & (3) & (4)  & (5) \\
 \hline
 1   & Redshift & $z$ & 0.0766 & 0.0766 & \\
 2   & Inclination angle & $i$ & 85 & 85 & deg \\
 3   & Galactic equivalent hydrogen column density  & $N^{\rm gal}_H$   & 1.4 & 1.4 & 10$^{20}$\,cm$^{-2}$  \\
 4   & Equivalent hydrogen column density of the host galaxy  & $N_{H, 1}$   & $6.8^{+1.3}_{-0.6}$ & $6.6^{+0.03}_{-0.03}$ & 10$^{20}$\,cm$^{-2}$  \\
 5   & Intrinsic line-of-sight equivalent hydrogen column density & $N_{H, 2}$ & $0.44^{+0.01}_{-0.02}$ & $0.32^{+0.09}_{-0.04}$ & 10$^{24}$\,cm$^{-2}$   \\
 6   & Photon index of the direct emission & $\Gamma$ & $1.45^{+0.11}_{-0.01}$ & $1.448^{+0.128}_{-0.002}$ & \\
 7   & Cut-off energy of the direct emission & $E_{\rm cut}$ & 300 & 300 & keV\\
 8   & Equivalent hydrogen column density of the torus  & $N_H^{\rm torus}$ & $N_{H, 2}$ & $1.3^{+0.9}_{-0.3}$  & 10$^{24}$\,cm$^{-2}$\\
 9   & Covering factor of the torus  & $\rm CF$  & $0.91^{+0.09}_{-0.32}$ & $0.91^{+0.09}_{-0.44}$ & \\
10   & Constant normalizing the line emission component  & $\rm C_{\rm line}$  & $3.5^{+0.4}_{-0.4}$ & $8.9^{+8.9}_{-2.5}$ & \\
11   & Iron abundance of the torus  & $\rm A_{Fe}$  & 1  & 1  & Solar\\
12   & Scattering fraction of the direct power law & $C_{\rm scat}$ & $0.61^{+0.05}_{-0.14}$ & $0.69^{+0.06}_{-0.18}$ & \\
13   & Cross-normalization constant & $A_{\rm chandra}$ & 1 & 1 & \\
14   & Cross-normalization constant & $A_{\rm xmm}$ & $1.05^{+0.03}_{-0.04}$ & $1.09^{+0.04}_{-0.03}$ & \\
15   & Cross-normalization constant & $A_{\rm nA}$  & $1.36^{+0.02}_{-0.05}$ & $1.38^{+0.18}_{-0.07}$ & \\
16   & Cross-normalization constant & $A_{\rm nB}$  & $1.28^{+0.07}_{-0.19}$ & $1.30^{+0.11}_{-0.02}$ & \\
\hline
17   & Soft 0.5--2\,keV observed frame flux$^b$  & $F_{\rm soft}$   &  $ 0.61 \pm 0.08$  & $0.62 \pm 0.08$
& $10^{-13}$\,erg\,cm$^{-2}$\,s$^{-1}$ \\
18   & Intermediate 2--10\,keV observed frame flux$^b$  & $F_{\rm intm}$   &  $2.9 \pm 0.4$ & $3.1 \pm 0.4$ & $10^{-13}$\,erg\,cm$^{-2}$\,s$^{-1}$ \\
19   & Hard 10--30\,keV observed frame flux$^b$  & $F_{\rm hard}$   & $6.5 \pm 0.8$  & $7.1 \pm 1.0$
& $10^{-13}$\,erg\,cm$^{-2}$\,s$^{-1}$ \\
20   & Soft 0.5--2\,keV rest-frame intrinsic luminosity$^b$  & $L_{\rm soft}$   & $ 1.7 \pm 0.2$ 
&  $1.5 \pm 0.2$ & $10^{42}$\,erg\,s$^{-1}$ \\
21   & Intermediate 2--10\,keV rest-frame intrinsic luminosity$^b$  & $L_{\rm intm}$   & $4.5 
\pm 0.6$  & $4.0 \pm 0.6$ & $10^{42}$\,erg\,s$^{-1}$ \\
22   & Hard 10--30\,keV rest-frame intrinsic luminosity$^b$  & $L_{\rm hard}$   & $6.1 \pm 
0.8$  & $5.5 \pm 0.7$ & $10^{42}$\,erg\,s$^{-1}$ \\
\hline
     & C-statistic & & 2308 & 2301 & \\
     & Degrees of freedom & & 2336 & 2335 & \\
\hline\hline
\end{tabular}
\end{center}
NOTES: $^a$ Columns (1) and (2) provide the description of the model parameters. Columns
(3) and (4) give model parameter values for the two scenarios, with $N_{H, 2}$ and $N_H^{\rm torus}$
linked to each other and unlinked, respectively. Column (5) gives the units of the model parameters.
$^b$ Mean and standard deviation of the 68\% confidence level measurements from the four
individual data sets.
}
\end{table*}

\section{Data and Spectral Modeling}
\label{sec:obs}

\subsection{\chandra}
\label{sec:chandra}

\oq\ was observed with \chandra\ ACIS-S for $\sim 30$\,ksec (ObsID=16071, see 
Table~\ref{tab:obs} for details). The target was placed at the aim point on the 
back-illuminated ACIS CCD (ACIS-S3). The observations were made in VFAINT mode with 
1/8 CCD readout (\chandra\ Proposers Observatory 
Guide\footnote{http://cxc.harvard.edu/proposer/POG/}).
The target was clearly detected with \chandra\ (see Fig.~\ref{fig:fig1}).

We used the \chandra\ CIAO software, version 4.9 (Fruscione et al. 2006) and CALDB
version 4.7.6 to process the data. We used CIAO tool {\tt acis-process-events} 
and applied the newest calibration files, filtered VFAINT background events, and ran 
a sub-pixel event-repositioning algorithm (and set {\tt pix\_adj=EDSER}). This final 
step provides the highest angular resolution X-ray image data for the most up-to-
date ACIS-S calibration. 
Figure~\ref{fig:fig1} displays the \chandra\ ACIS-S counts image in the 0.5--7\,keV 
energy range, with the central 1.5\arcsec\ (2.1\,kpc at the redshift of the source)
radius circle centered on the coordinates of \oq\ (14:07:00.41, +28:27:14.65)
marking the extraction region of the point source emission.
The entire X-ray emission is contained in this region with low background counts
scattered over the entire image. For the background, we used an 
annular region centered on the same position, with inner and outer radii of 
2$\arcsec$ and 5$\arcsec$.

On a larger scale we detected a point X-ray source located $30\arcsec$ ($\sim$44\,kpc
at $z=0.0766$) in the NE direction from the \oq\ center. Since the two radio lobes in
\oq\ extend along the NW-SW axis, it is not likely that this source could be associated
with a hot spot or past radio activity of the source. There is no obvious optical
counterpart to the secondary source in the Sloan Digital Sky Survey. We note that the
secondary, if persistent, would be included within the PN source extraction region in
Guainazzi et al. (2004). However, in our \chandra\ data we detect only 27 counts from
this source, as opposed to 650 counts from the \oq\ extraction region. Thus, we assess
that the secondary source does not contribute significantly to the \xmm\ spectrum of \oq.

\subsection{\nustar}
\label{sec:nustar}

The \nustar\ \citep{harrison2013} observation of \oq\ was performed on 2016-06-18 
for about 51\,ksec (ID 60201043002) using two telescopes, FPMA and FPMB. We used
HEAsoft 6.22.1 and {\tt nustardas\_06Jul17\_v1.8.0} software with the CALDB version
20171002 for data processing. We run {\tt nupipeline} and {\tt nuproduct} to
apply dead time correction, account for SAA passages with {\tt SAAMODE = 
optimize}, apply new calibration, and to extract spectra and generate appropriate 
response files. The source was clearly detected with \nustar\ (Fig.~\ref{fig:fig1}).
We assumed a circle with the default 49.2\arcsec\ radius centered on \oq\ 
coordinates (14:07:00.4, +28:27:14.6) for extracting the source spectra. 
For the background we used annular regions centered on the same position with inner
and outer radii of 123\arcsec\ and 196\arcsec, respectively.
Table~\ref{tab:obs} shows the resulting exposure and number of counts for each
telescope. 

\subsection{Spectral modeling}

Guainazzi et al. (2004) observed \oq\ with \xmm\ and inferred that it could be
a Compton thick AGN based on its apparently hard photon index and a strong neutral
iron line emission. However, \xmm\ data alone were not sufficient to constrain the 
photon index of the source and the amount of the intrinsic absorption.
Here, we model simultaneously the new \chandra\ and \nustar\ data, as well as the 
archival \xmm\ PN data of \oq\ using the {\tt XSPEC} 12.9.1 software package 
with C-stat statistic (Arnaud et al. 1996). We allow for cross-normalization
constants between the four data sets.

We construct a model in which the intrinsic X-ray emission of the source is 
described with a power-law function with photon index, $\Gamma$, and 
an exponential cutoff fixed at $E_{\rm cut} 
= 300$\,keV, absorbed by the Galactic and host galaxy hydrogen column densities
($N^{\rm gal}_H$ and $N_{H, 1}$, respectively). The 
X-ray emission above $\sim 4$\,keV contains a strong contribution from the 
reflection of the primary X-ray continuum from a cold matter, presumably a toroidal 
structure  around the central black hole, which we describe with the torus model of 
Balokovi\'c  et al. (2018). The level of the soft X-ray emission below 4\,keV 
requires that we include the scattered intrinsic continuum in the model.
The complete model can be written as: $\rm A_{instr} \times 
M_{\rm abs, 1} \times \left[ C_{\rm scat} \times {\tt cutoffpl} +  
M_{\rm abs, 2} \times {\tt cutoffpl} + {\tt torus} \right]$,
where $A_{\rm instr}$ stands for a constant introduced in order to
account for the uncertainties in the cross-normalization of the data sets from
the different telescopes; $M_{\rm abs, 1} = \exp \left(-N^{\rm gal}_H \sigma_E 
\right) \exp \left( -N_{H, 1} \sigma_{E(1 + z)} \right)$ and  
$\sigma_E$ is the photo-electric cross-section 
(in practice  $M_{\rm abs, 1}$ equals {\tt phabs * zphabs} in the {\tt XSPEC} 
nomenclature); $M_{\rm abs, 2} = \exp \left[ -N_{H, 2}(\theta) 
\sigma_{E(1+z)} \right]$ is the zeroth-order multiplicative table given by Yaqoob 
(2012) and it accounts for the angle dependent absorption of the primary continuum 
by the obscuring torus; $C_{\rm scat}$ is the normalization constant corresponding
to the scattered power-law emission. The torus component is described using
the table models provided by Balokovi\'c et al. (2018). They self-consistently compute 
the reflection continuum and fluorescent K$\alpha_1$, K$\alpha_2$ and K$\beta$
emission lines for atomic species up to zinc. Here,
${\tt torus} = {\tt ref\_cont} + C_{\rm line} \times {\tt lines}$, where we 
introduce a normalization constant $C_{\rm line}$ to account for the deviations from
the perfect torroidal geometry. We investigate two scenarios: 
one where the line-of-sight column density, $N_{H, 2}$, and the average torus column 
density, $N^{\rm torus}_H$, are linked to each other, and one where they are
allowed to vary independently (corresponding to a porous torus case).

Stanghellini et al. (1997a) reported that the milliarcsecond morphology of the
radio emission in \oq\ shows a flux density ratio between the two components
of about 10:1. Based on the optical image of the galaxy, they indicate that the jet
inclination of $\sim$\,45 deg is possible if the radio axis is aligned with the optical
axis. They argue that the bulk velocity of $\sim$\,0.5$c$ for this inclination angle
could explain this flux density ratio via Doppler boosting. However, they also point
out that the the difference in flux density may simply reflect random differences
between the paths of the jets. Here, we assume that the radio jet is oriented in the
plane of the sky (perpendicular to our line-of-sight), and the torus is perpendicular
to the jet axis. Thus, we fix the inclination angle at a value corresponding
to the $\sim 84-90$\, deg in the torus model of Balokovi\'c et al. (2018).

%%%%%%%%%%%% Figure 2

\begin{figure}
\includegraphics[width=0.95\columnwidth]{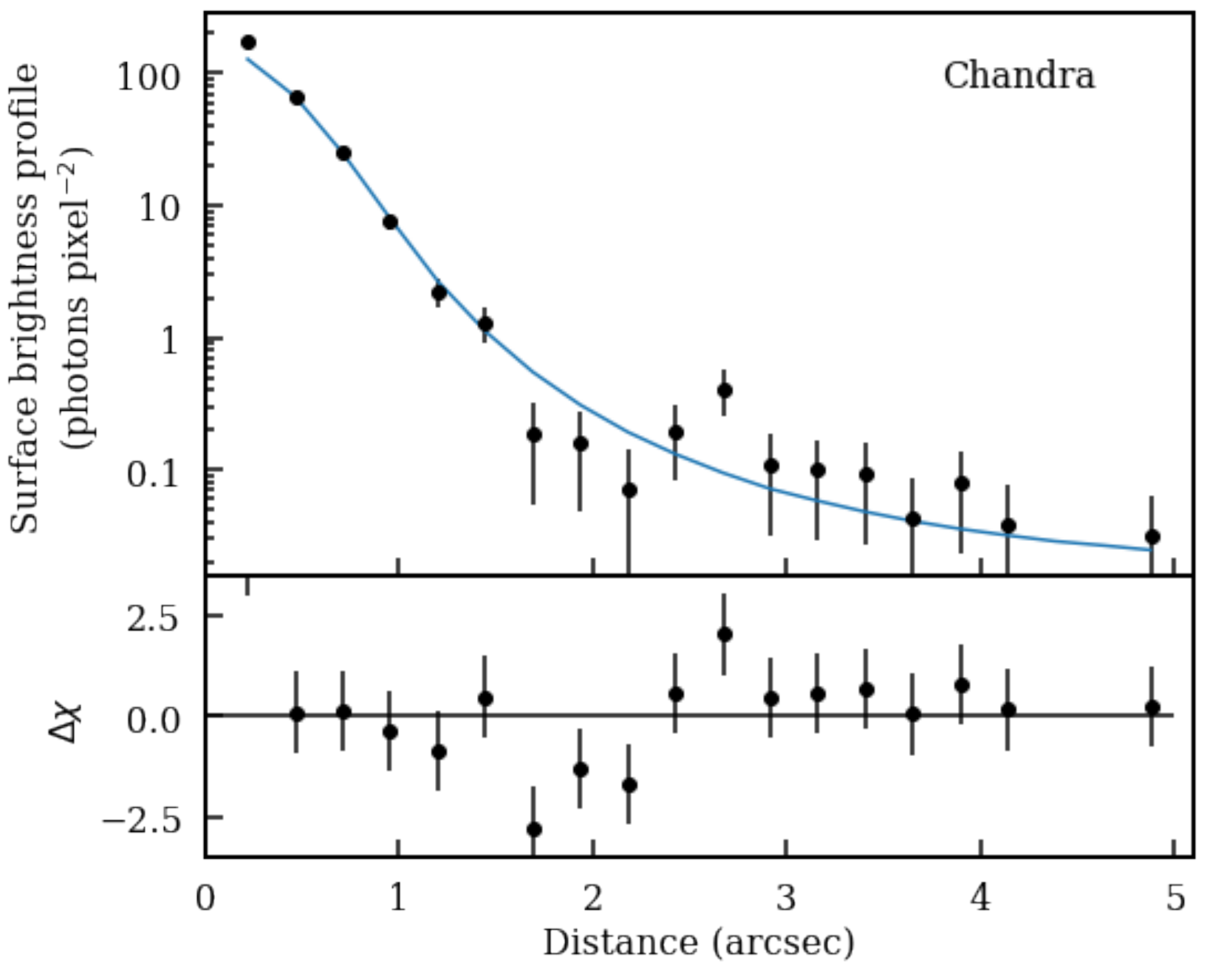}
\caption{{\bf Top:} Surface brightness profile of \oq\ resulting from the \chandra\ 
image. The solid line shows the model of \chandra\ point spread function
corresponding to a point source. {\bf Bottom:} Model residuals of the fit.
No evidence for an extended X-ray emission is detected.}
\label{fig:sb}
\end{figure}

\section{Results}
\label{sec:res}

\subsection{The first detection of hard X-ray emission from a Compact Symmetric Object}

Our \nustar\ observation of \oq\ provides the first hard X-ray ($> 10$\,keV)
detection of a CSO, and establishes these young radio sources as
a new class of hard X-ray emitters. The source is detected up to 30\,keV,
at the $6\sigma$ confidence level within the 10--30\,keV energy band.

\bigskip

\subsection{Broad-band X-ray spectrum of \oq}

%%%%%%%%%%%% Figure 3

\begin{figure}
\includegraphics[width=0.95\columnwidth]{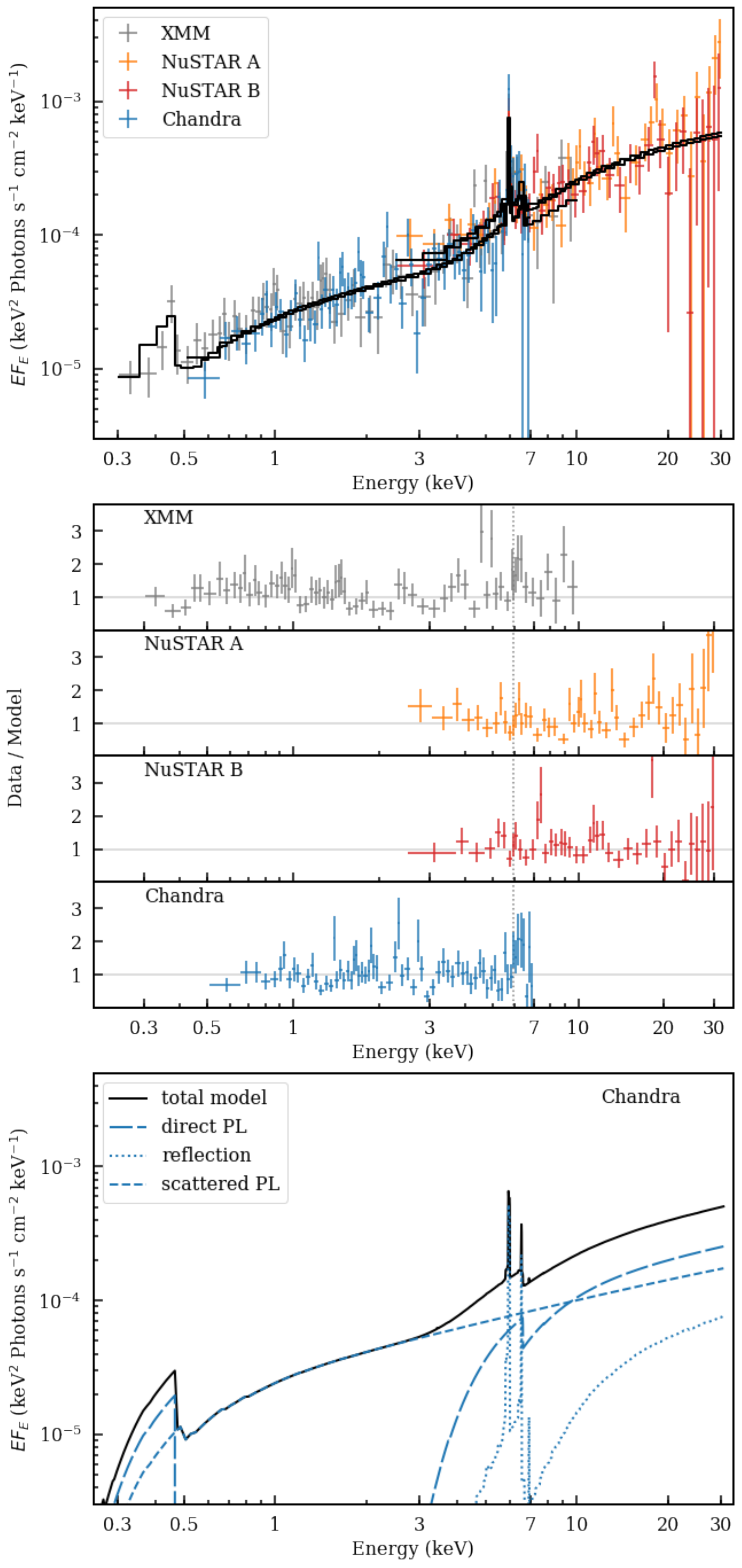}
\caption{{\bf Top:} Unfolded data and models from the simultaneous fit of the new 
\chandra\ and \nustar, and archival \xmm\ PN data sets in the scenario with the
line-of-sight and torus hydrogen column densities linked to each other.
The unlinked scenario results in a fit of comparable quality. {\bf Center:}
Corresponding data to model ratios. The vertical dotted line indicates the
rest-frame energy of the 6.4\,keV Fe fluorescent emission line.
{\bf Bottom:} Unfolded model (solid line) and model components
(direct absorbed power-law, long-dashed line;
reflection from a cold matter including fluorescent Fe emission, dotted line;
scattered power-law, short-dashed line)
computed over the 0.3--30\,keV energy band based on the model that provided the
best fit to the \chandra\ data.}
\label{fig:torus}
\end{figure}

We found that the spectral model described in Sec. 2.3 fits well 
our broad-band multi-instrument X-ray data. Both scenarios, with
linked and unlinked intrinsic equivalent hydrogen column densities
corresponding to the line-of-sight direction, $N_{H, 2}$, and the torus,
$N^{\rm torus}_H$, resulted in acceptable fits with C-statistic/degrees
of freedom of 2308/2336 and 2301/2335, respectively.
Neither the \chandra\ image nor the broad-band X-ray spectrum provided
evidence for the presence of an extended X-ray emission (Fig.~\ref{fig:sb}).
We show the data, unfolded models, model components, and data to model ratios for the 
case of linked column densities in Fig.~\ref{fig:torus}. They are representative
also of the unlinked column densities scenario.

We constrained the intrinsic photon index of the source, 
$\Gamma = 1.44^{+0.11}_{-0.01}$ and $\Gamma = 1.448^{+0.128}_{-0.002}$ for
the linked and unlinked case, respectively (Tab.~2). We derived the absorbing 
column density of the torus and the line-of-sight column density:
$N_{H, 2} = N^{\rm torus}_H$ and $3.2^{+0.9}_{-0.4} \times 10^{23}$\,cm$^{-2}$ (linked scenario),
and $N^{\rm torus}_H = 4.4^{+0.1}_{-0.2} \times 10^{23}$\,cm$^{-2}$ and $1.3^{+0.9}_{-0.3} \times 
10^{24}$\,cm$^{-2}$ (unlinked scenario).
This confirms the presence
of a significant amount of cold matter surrounding \oq\
with a large covering factor, $\rm CF > 0.5$--0.6, strongly attenuating
the primary X-ray continuum. The torus column density of the unlinked scenario
(porous torus) is $\sim 4$ times higher than in the linked scenario and high
enough to classify this source as a Compton thick AGN. Additionally, the data
require an intrinsic absorbing column of the order of
$N_{H, 1} \simeq 7 \times 10^{20}$\,cm$^{-2}$ attributable to the host galaxy.

We find evidence for a strong fluorescent Fe emission around 6\,keV
(observer's frame). With the iron 
abundance of the torus fixed at the solar value and $C_{\rm line} = 1$, our 
spectral model appeared to underestimate the strength of the Fe\,K$\alpha$
line emission. Thus, we allowed $C_{\rm line}$ to vary, and derived
$C_{\rm line} = 3.5^{+0.4}_{-0.4}$ and $8.9^{+8.9}_{-2.5}$ for the linked and
unlinked scenarios, respectively. Alternatively, we were able to obtain fits of
comparable quality by keeping $C_{\rm line} = 1$, and varying the iron abundance
of the torus, which resulted in abundance of $\sim 3$ and $\sim 9$ times the Solar
value for our two cases, accordingly. Allowing the normalization of the line
component to vary does not affect the derived values for the column densities.

The soft part of the observed spectrum, between 0.5 and 4\,keV, is dominated by
scattered power law emission with a relatively high scattering fraction of $C_{\rm 
scat} \simeq 0.6$--0.7.

Our modeling gives rest-frame 0.5--30\,keV de-absorbed luminosity
of \oq\ of the order of 10$^{43}$\,erg\,s$^{-1}$. Table~2 shows the contributions
to this luminosity from the soft (0.5--2\,keV), intermediate (2--10\,keV) and
hard (10--30\,keV) X-ray bands. We also list the observed fluxes (not corrected for
absorption) in these three energy bands (in the observed frame).

\subsection{X-ray variability}

We found that the \chandra\ and \xmm\ PN cross-normalization
constants are consistent with each other, indicating no flux variability
between the 2003 and 2014 observations. However, the
FPMA/B cross-normalization constants indicate that the normalization of
the \nustar\ spectra is higher than that of the \chandra\ spectrum
by up to $\sim 30$--40\% over the 0.5--30\,keV energy range used in spectral
fitting. According to Madsen et al. (2017), the \chandra\ calibration flux
rises by up to 10\%--15\% above 4\,keV compared to the \nustar\ flux.
Thus, it is likely that the X-ray flux of \oq\ varied between 2014 and
2016 by up to $\sim 50$\%.

\section{Discussion}
\label{sec:d}

Young radio sources have been proposed to be X-ray and $\gamma$-ray emitters
based on theoretical considerations in which their high energy emission
originates due to the inverse Compton scattering of ambient soft photon fields off
energetic electrons injected into expanding lobes from the hot spots (e.g. Stawarz
et al. 2008; Ostorero et al. 2010). A small sample of 16 CSOs has been observed
and detected in the soft X-rays with \chandra\ and/or \xmm\ (S16 and references 
therein),
and Migliori et al. (2016) reported the detection of PKS 1718+654 in the $\gamma$-ray
band with Fermi/LAT. Here, we presented new \chandra\ and \nustar\ observations of the 
young radio source \oq. The source was clearly detected with both instruments.
The \nustar\ detection constitutes the first hard X-ray ($>10$\,keV) detection of a CSO, 
and establishes CSOs as a new class of \nustar\ sources.
We modeled the \nustar\ and \chandra\ data simultaneously, including also the archival 
\xmm\ PN data set, assuming a model in which the direct power law X-ray
emission interacts with the obscuring cold matter surrounding the central black hole 
through absorption, reflection, and scattering processes (Balokovi\'c et al. 2018).

\subsection{Origin of the X-ray emission}

The broad-band X-ray coverage of our observations enabled us 
to constrain the photon index of the primary X-ray emission
from the source, and measure 0.5--30\,keV (rest-frame)
intrinsic luminosity of $L_{\rm X} \sim 10^{43}$\,erg\,s$^{-1}$.
The photon index of $\Gamma \simeq 1.45$ is relatively hard compared to
the photon indexes derived for \xmm\ and \nustar\ detected AGN
(e.g. Del Moro et al. 2017; Corral et al. 2011).
It is, however, rather typical of the X-ray spectra from accreting
black hole binaries in a hard jet-producing spectral state
(e.g. Done, Gierli\'nski \& Kubota 2007; Remillard \& McClintock 2006).
Power-law X-ray emission with such a photon index could be produced
e.g. in an accretion disk corona, expanding radio lobes,
or relativistic jets. However, the observed level of X-ray emission
is lower than the level of X-ray emission
due to the inverse Compton scattering of the infra-red and ultra-violet
photons off ultra-relativistic electrons in the radio lobes expected
for this source (Ostorero et al. 2010, their Fig. 1, Appendix A.9). This could
indicate that in the young radio lobes of \oq, the magnetic pressure dominates over
the electron pressure, rather than remaining in rough equipartition with
them as assumed in Stawarz et al. (2008). 
Detailed modeling
of the broadband radio-to-hard-X-ray SED is beyond the scope of this
work and will be addressed in the forthcoming publication.

Nonetheless, the bolometric luminosity of the source
derived in W\'ojtowicz et al. (2019, in preparation)
using the Richards et al. (2006) bolometric correction
applied to the observed 12\,$\mu$m luminosity (Kosmaczewski et al. 2019)
yields $L_{\rm bol} \simeq 4.3 \times 10^{45}$\,erg\,s$^{-1}$.
For the black hole mass of $M_{\rm BH} = 5 \times 10^8$\,M$_{\odot}$ (Wu 2009), and 
$L_{\rm Edd} = 1.3 \times 10^{38} (M_{\rm BH} / M_{\odot})$, this
bolometric luminosity corresponds to the Eddington luminosity ratio of
$\lambda_{\rm Edd} = L_{\rm bol}/L_{\rm Edd} \simeq 0.06$.
However, the 2--10\,keV X-ray luminosity expected from a radio loud
AGN with a similar bolometric luminosity
is of the order of $(3 - 7) \times 10^{43}$\,erg\,s$^{-1}$,
based on the bolometric corrections obtained by Runnoe et al. (2012),
i.e. an order of magnitude higher than the 2-10\,keV luminosity resulting from
our modeling of the X-ray spectrum of \oq\ (see Table 2).
It is thus possible that the X-ray radiation mechanism
of OQ+208 is less efficient than that of the quasars used by Runnoe et al. to
derive the bolometric corrections, or the bolometric luminosity
based on the 12\,$\mu$m luminosity is overestimated.

\subsection{Properties of the X-ray absorber}

Our observations allowed us to gain important insights into the
nature of the intrinsic obscuration in \oq. We adopted
the torus model of Balokovi\'c et al. (2018)
and detected intrinsic absorption
with the equivalent hydrogen column density of the order of 
10$^{23}$--10$^{24}$\,cm$^{-2}$ due to cold matter
surrounding the central black hole with a large covering
factor. It seems plausible that the torus has a
porous structure, given the good fit to the data of the model with
the line-of-sight absorbing column and torus absorbing column
varying freely and resulting in values that are statistically
different from each other. This conclusion is further supported by the
detection of a significant amount of the scattered primary
emission dominating the soft X-ray spectrum below 4\,keV.
With a porous structure, the primary emission would undergo less
attenuation as it could arrive to the observer through the holes
in the torus. The optical classification as a broad-line QSO
is also consistent with the porous structure of the torus obscuring
an X-ray source seen from a direction typical to Type 2 QSO.
Additionally, variable degree of the porosity of the obscurer
could explain the $\sim 50$\% X-ray flux variability
that the source appeared to undergo between our \chandra\ and
\nustar\ observations.

Alternatively, \oq\ could belong to the so-called type-12 AGN class,
optically un-obscured but X-ray obscured,
in which the X-ray obscuration was argued to be produced by dust-free gas
within (or inside) the broad line region (e.g. Merloni et al. 2014,
and references therein). However, it was shown that
majority of such AGN tend to have rather high X-ray luminosities
compared to \oq\ 
(e.g. Merloni et al. 2014; Marchesi et al. 2016).

When modeling only the \xmm\ data, Guainazzi et al. (2004)
found that the emission feature around 6\,keV (observer's frame) can
be modeled either as a single broad line with energy corresponding to
the neutral Fe, or by a combination of three
narrow lines, whose energies are consistent with neutral,
He-like, and H-like Fe. Our modeling corresponds to the latter scenario
(Balokovi\'c et al. 2018).

Moreover, we observed that the fluorescent iron line emission was enhanced 
relative to the strength of the associated Compton reflection continuum.
This may indicate a significant departure of the Fe-K reflection region from
the structure of our model,
or alternatively suggest elevated iron abundance in the matter
forming the torus. Importantly, 
6.4 keV line emission from an extended hundred pc-scale
region has been resolved in a few nearby galaxies (e.g. Fabbiano et al. 2018, 2019).
Additionally, an enhanced line emission with spatially variable EW has been reported
by Marrinucci et al. (2017).
The spatial scales studied by these authors cannot be resolved in X-rays in
our source,
nor is the possible extended 6.4 keV emission accounted for by the existing torus models.

We did not find evidence for a diffuse X-ray component in \oq,
contrary to the work of Lanz et al. (2016) based on the
\chandra\ data alone. In fact, the combined \chandra, \xmm, and
\nustar\ observations show that their model (two thermal components
and an absorbed power law) cannot account for the 6-7\,keV Fe emission
signature and the overall shape of the hard X-ray spectrum.

\section{Conclusions}
\label{sec:c}

Our results indicate that in \oq\ the young $\sim 250$
year old radio lobes spanning the length of 10\,pc coexist with
a cold obscuring matter, possibly a dusty torus. The structure of the torus
is likely porous, with high equivalent column density comparable to
that detected in Compton thick AGN. The X-ray source is powered by accretion
with a relatively high Eddington ratio of $\lambda_{\rm Edd} \sim 0.06$,
and emits a rather hard power-law
spectrum which might have increased in flux by $\sim 50$\% between our 2014
and 2016 observations. The X-ray emission is consistent with a point source
based on the \chandra\ image, and it is not likely to originate from the 
young radio lobes, unless the lobes in \oq\ are dominated by the magnetic
pressure.

\acknowledgements

This research has made use of data obtained by the \chandra\ X-ray Observatory and software 
provided by the Chandra X-ray Center (CXC). It made also use of data from the NuSTAR mission, 
a project led by the California Institute of Technology, managed by the Jet Propulsion 
Laboratory, and funded by the National Aeronautics and Space Administration. This research has 
made use of the NuSTAR Data Analysis Software (NuSTARDAS) jointly developed by the ASI Science 
Data Center (ASDC, Italy) and the California Institute of Technology (USA). This project
was supported in part by the NASA grants GO4-15099X and NNX17AC23G.
M.S. and A.S. were supported by NASA contract NAS8-03060 (Chandra X-ray Center).
M.S. acknowledges the Polish NCN grant OPUS 2014/13/B/ST9/00570. 
\L{}.S. was supported by Polish NSC grant 2016/22/E/ST9/00061.

\end{document}